\documentclass[pdflatex,sn-mathphys-num]{sn-jnl}

\usepackage{graphicx}%
\usepackage{multirow}%
\usepackage{amsmath,amssymb,amsfonts}%
\usepackage{amsthm}%
\usepackage{mathrsfs}%
\usepackage[title]{appendix}%
\usepackage{xcolor}%
\usepackage{textcomp}%
\usepackage{manyfoot}%
\usepackage{booktabs}%
\usepackage{array}%
\usepackage{algorithm}%
\usepackage{algorithmicx}%
\usepackage{algpseudocode}%
\usepackage{listings}%
\usepackage{booktabs}%
\usepackage{float}%

\theoremstyle{thmstyleone}%
\theoremstyle{thmstyletwo}%

\theoremstyle{thmstylethree}%

\begin{document}

\title[Article Title]{AutoAssert 1: A LoRA  Fine-Tuned LLM Model for Efficient Automated Assertion Generation}


\author*[1,2]{\sur{Yi Zhong}}\email{zhongyi@bupt.edu.cn}

\author[1]{\sur{Hongchao Liu}}
\author[1,3]{\sur{Di Zhao}}

\affil[1]{\orgdiv{Institute of Computing Technology}, \orgname{Chinese Academy of Sciences}, \orgaddress{\street{No. 6, South Academy Road}, \city{Beijing}, \postcode{	100190}, \state{Haidian}, \country{China}}}
\affil[2]{\orgdiv{International College}, \orgname{Beijing University of Posts and Telecommunications}, \orgaddress{\street{No.10, Xitucheng Road}, \city{Beijing}, \postcode{100088}, \state{Haidian}, \country{China}}}
\affil[3]{\orgname{University of Chinese Academy of Sciences}, \orgaddress{\street{No. 80, Zhongguancun East Road}, \city{Beijing}, \postcode{100190}, \state{Haidian}, \country{China}}}


\abstract{As the complexity of software systems continues to increase, the demand for automated testing and maintenance tools is growing exponentially. To meet this urgent need, we propose a new assertion generation method based on Hardware Description Language (HDL). This method combines a lightweight, parameter-adjustable large language model (LLM) with the Unsloth platform to automatically generate test cases, thereby significantly reducing training costs without sacrificing accuracy or generalization performance.
Empirical evaluation shows that our method can efficiently generate assertions that strictly conform to the hardware logic. This framework provides a robust and flexible solution to modern software testing and maintenance challenges. https://github.com/liusu-orange/AutoAssert-1 and https://gitee.com/OpenBPU/auto-assert1 are the locations of the source code.}

\keywords{Assertion generation, Large Language Model, Fine-tuning, Hardware description language, Automated testing}



\maketitle

\section{Introduction}\label{sec1}
With the rapid development of hardware systems, their complexity has grown exponentially, making manual testing and maintenance both time-consuming and error-prone. Traditional automated test generation methods in hardware verification \cite{foster2016towards,spear2008systemverilog} often rely on rigid rule-based systems or manual scripting, lacking the flexibility to handle modern design complexities.
For example, in the verification of multi-core processors, the combinatorial explosion of states makes exhaustive testing impossible, often leaving critical edge cases uncovered, only 60\% to 70\% of edge cases are covered. \cite{gupta1992formal} Another notable case is the Intel Pentium FDIV bug, where a flaw in the floating-point division unit escaped detection due to incomplete test coverage, costing the company over \$475 million in recalls. 

To address these issues, the progress of large language models (LLMs) \cite{brown2020language,chen2021evaluating,an2024make} has shown great potential in automating complex tasks such as code generation and test case synthesis. However, fine-tuning these models for specific domain tasks (e.g., generating assertions from hardware description languages) faces several critical challenges in hardware verification:

Firstly, the validation system for training large models mainly relies on specific domain datasets, such as annotated RTL code or SystemVerilog assertions~\cite{lu2024rtllm, menon2025enhancing}. However, creating such datasets requires a significant amount of human effort and deep professional knowledge, which often leads to a scarcity of high-quality training samples. 

Secondly, general-purpose pre-trained language models often perform poorly when applied to hardware-related tasks because their understanding of hardware-specific grammar and semantics is limited. Although they perform well in natural language processing, they usually have difficulty simulating the inherent time constraints and concurrent characteristics of hardware systems, resulting in grammatically invalid or logically incorrect generated outputs.

Thirdly, fine-tuning to apply large language models to the hardware domain requires a large amount of computing resources, such as distributed frameworks and high-performance GPUs. Due to the need to frequently update model parameters based on evolving architectures (such as RISC-V, ARMv9), this increases the overall training cost. For example, training a model with 7 billion parameters may require an 8-fold A100 GPU cluster to run for 48 hours, with a cost of approximately \$5000.

This paper presents a novel approach that applies the pre-trained large-scale language model to the assertion generation task in hardware testing by minimizing parameter updates \cite{howard2018universal}. Our method utilizes the Unsloth platform \cite{rajbhandari2020zero}, significantly reducing the computational cost while maintaining high accuracy and generalization ability. In addition, we also adopt the lightweight Lora fine-tuning strategy \cite{hu2022lora,lester2021power} to freeze the original parameter model and only update a small number of Lora parameters. This method not only reduces the cost of model adaptation but also provides a new idea for deploying large-scale language models in resource-constrained environments.
The contributions of this article are as follows:
\begin{itemize}
    \item  We introduce a lightweight methodology that achieves competitive performance with minimal parameter updates, making it feasible for practical deployment.
    
    \item  Our work specifically targets assertion generation for hardware testing, addressing a critical gap in automated verification tools.
    
    \item  Through extensive experiments, we demonstrate that our approach outperforms traditional rule-based methods and reduces the computational burden compared to full-model fine-tuning.
\end{itemize}

\section{Related work}
The development of assertion generation techniques in hardware verification has undergone an evolution process from traditional rule-based methods to modern data-driven methods. Early research mainly relied on formal languages and manual templates, such as the Attribute Description Language and SystemVerilog Assertions (SVA) \cite{bergeron2012writing}\cite{spear2008systemverilog}, which provided standardized assertion syntax specifications. However, these methods required designers to have professional hardware knowledge and were difficult to adapt to new design requirements. Template-based tools like FoCs developed by IBM could automatically generate some assertions, but still had issues of insufficient flexibility and requiring a lot of manual intervention \cite{foster2016towards}.

With the development of machine learning technology, researchers began to explore data-driven assertion generation methods \cite{pulavarthi2024assertionbench}. In supervised learning \cite{iman2025advanced,yu2022automated}, there were studies that used support vector machines to analyze simulation trajectories to predict assertion templates, while works like DeepAssert used LSTM networks to generate assertions from HDL code  \cite{xue2009svm,wang2024deep}. However, these methods generally faced difficulties in obtaining labeled data. Reinforcement learning methods such as RLAssert modeled assertion generation as a reinforcement learning problem, although it reduced manual intervention, due to the need to design complex reward functions, it also had high computational costs.

In recent years, large language models \cite{brown2020language}\cite{hu2022lora} have demonstrated strong capabilities in code generation tasks, providing new solutions for assertion generation. General large models like GPT-3 \cite{floridi2020gpt} can generate code snippets similar to assertions based on natural language prompts, but they have the problem of insufficient domain specificity and low prediction accuracy for specific hardware description languages. VeriGen \cite{thakur2024verigen} and other studies adapted pre-trained models to the SystemVerilog assertion generation task through full parameter fine-tuning, although the results were significant, the computational resource consumption was huge. Emerging efficient fine-tuning platform like Unsloth \cite{rajbhandari2020zero}\cite{zheng2403unified}, although they could significantly reduce computational costs, have not yet been fully explored in the field of hardware verification. 

In summary, existing methods still have significant deficiencies in domain adaptability and computational efficiency, which provide an innovative space for this research. Our proposed lightweight supervised fine-tuning method \cite{hu2022lora}\cite{howard2018universal} not only significantly reduces the parameter update volume but also improves the model's professionalism through domain adaptation, ensuring performance while significantly reducing computational costs. Through systematic experiments, this method shows significant advantages over traditional rule-based methods and full parameter fine-tuning, providing an efficient and practical new approach for hardware assertion generation.

\section{Framework and Method}
\subsection{Framework Overview}
\begin{figure}[htbp]
\centering
\includegraphics[width=\linewidth]{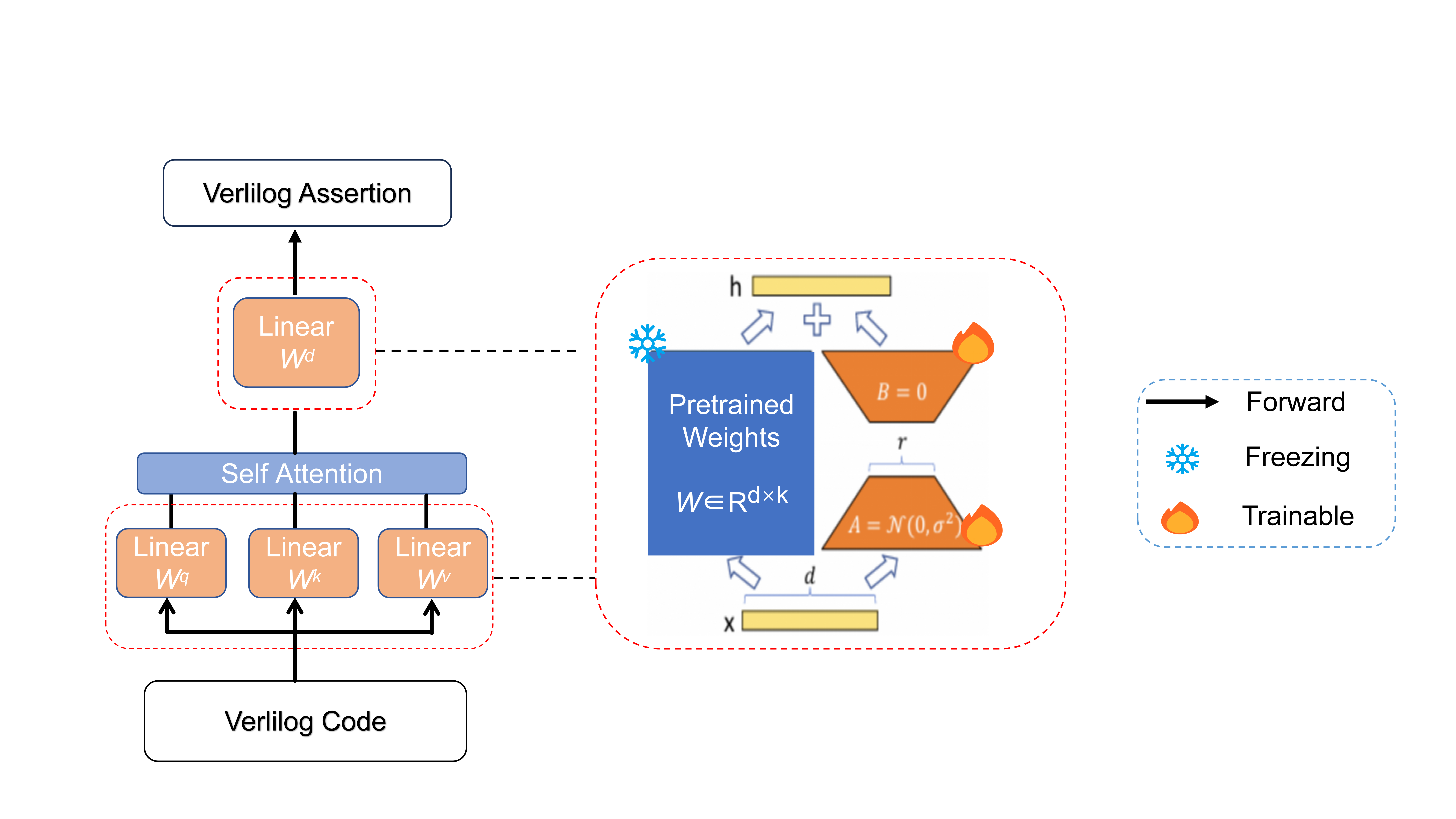}
\caption{This framework achieves the best balance between computational efficiency and model performance by keeping the parameters of the base model frozen while only making low-rank adaptive adjustments to the key projection layers in the attention mechanism and the feedforward network.}
\label{fig:lora_arch}
\end{figure}

The hardware assertion generation framework proposed in this study is based on the Unsloth efficient fine-tuning platform and adopts the LoRA paradigm to adapt large language models to specific domains \cite{chen2021evaluating}\cite{howard2018universal}. You can see the architecture in Fig. \ref{fig:lora_arch}.This framework maintains the parameter freezing of the base model while only performing low-rank adaptation on the key projection layers of the attention mechanism and the feedforward network, achieving the optimal balance between computational efficiency and model performance.
\subsection{LoRA Fine-Tuning Principle}
LoRA \cite{hu2022lora} (Low-Rank Adaptation) is a parameter-efficient fine-tuning method that approximates full parameter updates through low-rank decomposition, as illustrated in Fig. \ref{fig:lora_arch}. 
Given an original weight matrix \( W \in \mathbb{R}^{d \times k} \) in a pre-trained model, LoRA introduces two low-rank matrices: \( A \in \mathbb{R}^{d \times r} \) and \( B \in \mathbb{R}^{r \times k} \), where the rank \( r \) satisfies \( r \ll \min(d,k) \). The matrix \( A \) is randomly initialized, while \( B \) is initialized to zero. The parameter update \( \Delta W \) is computed as  shown in the equation (\ref{eq1}) and equation (\ref{eq2}).
\begin{equation}
\Delta W = BA
\label{eq1}
\end{equation}

The fine-tuned weight matrix \( W' \) is then given by:
\begin{equation}
W' = W + \Delta W = W + BA
\label{eq2}
\end{equation}

The rank \( r \) is typically much smaller than the original dimensions \( d \) and \( k \) (e.g., \( r = 8 \) for \( d,k \geq 1024 \)). This reduces the number of trainable parameters from \( d \times k \) to \( r \times (d + k) \), significantly lowering memory and computational costs. 

During backpropagation, gradients flow only through \( B \) and \( A \), leaving the original weights \( W \) frozen. This prevents catastrophic forgetting of pre-trained knowledge while adapting to new tasks. 

The matrix \( A \) is initialized with random Gaussian noise scaled by \( 1/\sqrt{r} \), while \( B \) is zero-initialized to ensure \( \Delta W = 0 \) at the start of training. This stabilizes early fine-tuning.

In our study, we apply LoRA to critical projection layers in the Transformer architecture, targeting the attention layers (\texttt{q\_proj}, \texttt{k\_proj}, \texttt{v\_proj}, and \texttt{o\_proj}) and feed-forward network layers (\texttt{gate\_proj}, \texttt{up\_proj}, and \texttt{down\_proj}). This layer selection strategy is theoretically grounded in the observation that attention layers directly determine token interaction patterns crucial for semantic understanding, while feed-forward network projections govern nonlinear feature transformations essential for domain knowledge encoding. By setting the rank $r=16$, we reduce the trainable parameters in each target layer to less than 0.1\% of the original parameters. For typical dimensions of $d=4096$ and $k=4096$, this yields a compression ratio of approximately $0.52\%$, demonstrating the remarkable efficiency of our approach while maintaining model performance.

\section{Dataset and Experiments}
\subsection{Introduction of dataset}
This study uses the VERT dataset \cite{menon2025enhancing} to evaluate and train the hardware assertion generation task. You can find this dataset at https://github.com/AnandMenon12/VERT. The dataset contains 20,000 pairs of carefully annotated Verilog/SystemVerilog code and assertions, and it is divided into various categories to ensure coverage of all scenarios in the hardware domain. Among them, 460 categories are derived from the Xiangshan processor.

In the experiment, we selected 18,000 pairs of data from the dataset for fine-tuning the model as the training set, 1,000 pairs of data for adjusting the hyperparameters as the validation set, and another 1,000 pairs of data for the final performance evaluation as the test set. All assertions in the dataset strictly follow the SystemVerilog assertion (SVA) syntax standard and cover complete verification scenarios ranging from basic logic constraints to timing requirements.
\subsection{Experimental Design}

\subsubsection{Pretrained Model Comparison}
As shown in Table~\ref{tab:model_compare}, we evaluate four open-source LLMs in the context of hardware assertion generation. Among them, LLaMA-3-7B~\cite{touvron2023llama} achieves the best performance, primarily due to its English-centric pretraining.In contrast, Qwen-7B~\cite{bai2023qwen}, with its multilingual orientation and Chinese-focused training corpus, exhibits weaker compatibility with HDL syntax despite its robust handling of extended contexts. DeepSeek-LLM-7B~\cite{liu2024deepseek}, designed for Chinese-English code blending, introduces unnecessary complexity when applied to monolingual hardware domains. Similarly, although Gemma-7B~\cite{team2024gemma} features low-latency optimizations, this comes at the cost of reduced accuracy in generating temporal logic assertions. These findings suggest that for HDL-related tasks, linguistic precision and syntactic fidelity are more critical than multilingual capabilities or inference speed.

In this comparison, we use five commonly adopted evaluation metrics for text generation tasks: BLEU, ROUGE-1, ROUGE-2, ROUGE-L, and Accuracy. Below are their definitions and corresponding formulas.

BLEU (Bilingual Evaluation Understudy) measures the similarity between generated text and reference text. BLEU calculates the precision of multiple n-grams and includes a brevity penalty (BP) to penalize short outputs. The BLEU score is computed as shown in equation(\ref{eq:bleu}):

\begin{equation}
\text{BLEU} = \text{BP} \cdot \exp\left(\sum_{n=1}^{N} w_n \log p_n \right)
\label{eq:bleu}
\end{equation}

Here, \( p_n \) is the precision of n-grams, \( w_n \) is the weight (usually uniform),\( c \) is the length of the candidate sentence and \( r \) is the length of the reference sentence and BP is the brevity penalty defined in equation(\ref{eq:bp}):

\begin{equation}
\text{BP} = 
\begin{cases}
1 & \text{if } c > r \\
e^{1 - \frac{r}{c}} & \text{if } c \leq r
\end{cases}
\label{eq:bp}
\end{equation}
where \( c \) is the length of the candidate sentence and \( r \) is the length of the reference sentence.

ROUGE-N evaluates the recall of n-grams between the generated and reference texts. ROUGE-1 computes unigram recall, while ROUGE-2 computes bigram recall. The ROUGE-N score is given in equation(\ref{eq:rouge-n}):

\begin{equation}
\text{ROUGE-N} = \frac{\sum \text{overlap}_n}{\sum \text{reference}_n}
\label{eq:rouge-n}
\end{equation}

ROUGE-L is based on the Longest Common Subsequence (LCS), capturing sentence-level structure similarity. The F-score version of ROUGE-L is defined in equation(\ref{eq:rouge-l}):

\begin{equation}
\text{ROUGE-L} = \frac{(1 + \beta^2) \cdot R \cdot P}{R + \beta^2 \cdot P}
\label{eq:rouge-l}
\end{equation}
where the precision \( P \) and recall \( R \) are calculated as shown in equation(\ref{eq:pr}):

\begin{equation}
P = \frac{\text{LCS}}{\text{Candidate Length}}, \quad R = \frac{\text{LCS}}{\text{Reference Length}}
\label{eq:pr}
\end{equation}

Accuracy reflects the proportion of correct predictions among all predictions. The accuracy score is given in equation(\ref{eq:acc}):

\begin{equation}
\text{Accuracy} = \frac{\text{Correct Predictions}}{\text{Total Predictions}}
\label{eq:acc}
\end{equation}

\begin{table}[htbp]
\centering
\caption{Performance comparison of different pretrained models based on BLEU, ROUGE-1, ROUGE-2, and ROUGE-L metrics}
\begin{tabular}{lccccc}
\toprule
\textbf{Model} & \textbf{BLEU} & \textbf{ROUGE-1} & \textbf{ROUGE-2} & \textbf{ROUGE-L} & \textbf{Accuracy} \\
\midrule
Qwen-7B &0.83  & 0.81 & 0.82 & 0.85 & 0.96 \\
DeepSeek-LLM-7B & 0.81 & 0.86 & 0.82 & 0.84 & 0.95 \\
Gemma-7B & 0.82 & 0.84 & 0.81 & 0.84 & 0.96 \\
LLaMA-3-7B & 0.83 & 0.87 & 0.83 & 0.86 & 0.97 \\
\bottomrule
\end{tabular}
\label{tab:model_compare}
\end{table}
\subsubsection{Comparison of Closed-Source Large Models}
As shown in Table \ref{tab:close-model_compare}, we analyzed four of the current mainstream closed-source large models \cite{team2023gemini} \cite{brown2020language}. The experimental results show that these models have demonstrated outstanding performance in the assertion generation task. However, there are also some issues, such as the opacity of the generation process and the high operational costs. Therefore, semiconductor companies are increasingly inclined to use open-source models for fine-tuning and local deployment. This preference mainly stems from the excellent controllability provided by the open-source framework, which can perform deep architecture optimization based on specific hardware configurations and performance requirements.
\begin{table}[htbp]
\centering
\caption{Performance comparison of different  Closed-Source Large  models based on BLEU, ROUGE-1, ROUGE-2, and ROUGE-L metrics}
\begin{tabular}{lccccc}
\toprule
\textbf{Model} & \textbf{BLEU} & \textbf{ROUGE-1} & \textbf{ROUGE-2} & \textbf{ROUGE-L} & \textbf{Accuracy} \\
\midrule
GPT 4 &0.82  & 0.81 & 0.82 & 0.87 & 0.96 \\
Grok 3 & 0.81 & 0.83 & 0.80 & 0.86 & 0.96 \\
Gemimi 2.5 Flash & 0.81 & 0.84 & 0.83 & 0.82 & 0.95 \\
Claude Sonnet4 & 0.81 & 0.86 & 0.83 & 0.84 & 0.98 \\
\bottomrule
\end{tabular}
\label{tab:close-model_compare}
\end{table}
\subsubsection{LoRA Hyperparameter Study}

As shown in Table \ref{tab:r_ablation} and Table \ref{tab:r_ablation2}, we evaluate LoRA rank ($r$) configurations ($r=8,16,32$) on LLaMA-3-7B with fixed hyperparameters ($\alpha=16$, dropout = 0). Performance improves with higher ranks (97\% accuracy at $r=16$ vs. 91\% accuracy at $r=8$), but exhibits diminishing returns: the gain from $r=16$ to $r=32$ is significantly smaller than from $r=8$ to $r=16$, while requiring double the training parameters(41M → 83M). This suggests $r=16$ achieves the optimal balance, as the marginal improvement at $r=32$ does not justify its computational overhead for hardware assertion generation tasks.

As shown in Table \ref{tab:alpha}. We evaluate the impact of LoRA alpha values (8, 16, 32) on model performance. The results demonstrate that alpha=16 consistently outperforms the other values (alpha=8 and alpha=32) in accuracy, BLEU, and ROUGE metrics. Specifically, alpha=8 may limit the model's learning capacity due to its small size, while alpha=32 could lead to overfitting or instability. Alpha=16 strikes an optimal balance between learning and generalization.
\begin{table}[h]
\centering
\caption{Performance comparison of different rank (r) settings based on BLEU, ROUGE-1, ROUGE-2, and ROUGE-L metrics}
\label{tab:r_ablation}
\begin{tabular*}{0.85\textwidth}{@{\extracolsep{\fill}}cccccc}
\toprule
\textbf{R value} & \textbf{BLEU} & \textbf{ROUGE-1} & \textbf{ROUGE-2} & \textbf{ROUGE-L} & \textbf{Accuracy} \\
\midrule
8 & 0.79 & 0.81 & 0.77 & 0.81 & 0.91 \\
16 & 0.83 & 0.87 & 0.83 & 0.86 & 0.97 \\
32 & 0.84 & 0.87 & 0.85 & 0.88 & 0.97 \\
\bottomrule
\end{tabular*}
\end{table}

\begin{table}[h]
\centering
\caption{Comparison of the number of training parameters and running time under different rank settings}
\label{tab:r_ablation2}
\begin{tabular*}{0.85\textwidth}{@{\extracolsep{\fill}}cccc}
\toprule
\textbf{R value} & \textbf{Number} & \textbf{Percentage} & \textbf{Running time} \\
\midrule
8  & 21M & 0.26\% & 95 min \\
16 & 41M & 0.52\% & 175 min \\
32 & 83M & 1.10\% & 315 min \\
\bottomrule
\end{tabular*}
\end{table}

\begin{table}[h]
\centering
\caption{Performance comparison of different alpha settings based on BLEU, ROUGE-1, ROUGE-2, and ROUGE-L metrics}
\begin{tabular*}{0.85\textwidth}{@{\extracolsep{\fill}}cccccc}
\toprule
\textbf{Alpha value} & \textbf{BLEU} & \textbf{ROUGE-1} & \textbf{ROUGE-2} & \textbf{ROUGE-L} & \textbf{Accuracy} \\
\midrule
8 & 0.78 & 0.80 & 0.76 & 0.81 & 0.89 \\
16 & 0.83 & 0.87 & 0.83 & 0.86 & 0.97 \\
32 & 0.81 & 0.84 & 0.81 & 0.82 & 0.93 \\
\bottomrule
\end{tabular*}
\label{tab:alpha}
\end{table}

As shown in Table~\ref{tab:module_ablation}, we systematically evaluate three LoRA module configurations with fixed rank $r=16$: (1) \textbf{Attention Layers} (\texttt{q\_proj}, \texttt{k\_proj}, \texttt{v\_proj}, \texttt{o\_proj}), which govern token-to-token interactions through query-key-value transformations and attention head fusion; (2) \textbf{FFN Layers} (\texttt{gate\_proj}, \texttt{up\_proj}, \texttt{down\_proj}), responsible for nonlinear feature space transformations via gating and dimensional scaling; and (3) \textbf{All Layers}, combining both groups for joint adaptation. 

The experimental results demonstrate that applying LoRA to \textbf{All Layers} yields the best performance. The combined configuration achieves a 4.3\% accuracy improvement over attention-only adaptation and 6.5\% accuracy over FFN-only adaptation, with only a 0.18\% increase in trainable parameters compared to single-module configurations. This indicates that in the task of generating hardware assertions, it is crucial to fully adapt to the architectural components, as no single type of module alone can fully capture the complex semantics of the hardware description language.

\begin{table}[htbp]
\centering
\caption{Performance comparison of different target module groups based on BLEU,
ROUGE-1, ROUGE-2, and ROUGE-L metrics}
\begin{tabular}{lccccc}
\toprule
\textbf{Module Group} & \textbf{BLEU} & \textbf{ROUGE-1} & \textbf{ROUGE-2} & \textbf{ROUGE-L} & \textbf{Accuracy} \\
\midrule
Attention Layers & 0.79 & 0.86 & 0.77 & 0.81 & 0.93 \\
FFN Layers & 0.77 & 0.84 & 0.75 & 0.82 & 0.91 \\
All Layers & 0.83 & 0.87 & 0.83 & 0.86 & 0.97 \\
\bottomrule
\end{tabular}
\label{tab:module_ablation}
\end{table}
\subsubsection{Fine-tuning platform}
In our training process, we adopted the Unsloth platform, which has many advantages compared to traditional model fine-tuning frameworks such as Hugging Face Transformers and PyTorch \cite{paszke2019pytorch}. Although these mature platforms have become the default standard for fine-tuning large language models, they often have some persistent issues, including high GPU memory consumption, low computational efficiency, and unstable performance during the Low-Rank Adaptation (LoRA) process.

Unsloth directly addresses these challenges through three pioneering methods. Firstly, it uses 4-bit quantization technology to store the weights of the base model, reducing GPU memory usage by approximately 70\%, effectively alleviating the memory bottleneck problem. Secondly, the platform integrates fusion kernel operations, optimizing the efficiency of core computing tasks. Finally, Unsloth introduces gradient clipping and layer normalization calibration mechanisms specifically for dealing with the numerical instability issues commonly encountered in LoRA training. These measures collectively reduce the risk of gradient explosion and overflow, ensuring a more stable and reliable training process.

\subsubsection{Training Details}
\begin{figure}
    \centering
    \includegraphics[width=\linewidth]{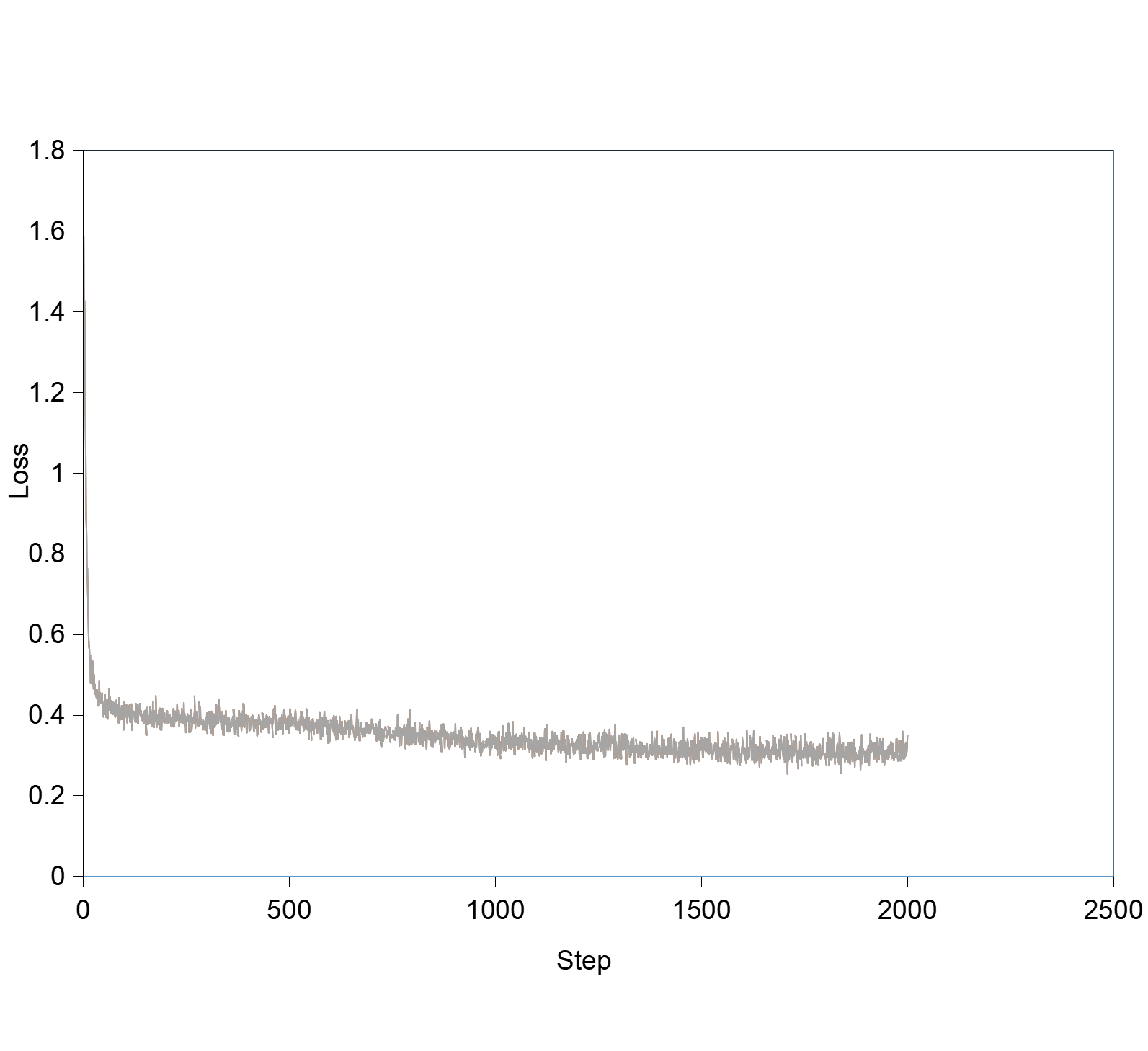}
    \caption{As can be seen from the graph, in the early stage of the iteration, the loss value drops rapidly, and then gradually stabilizes. Overall, it shows a trend of first decreasing rapidly and then remaining stable, indicating that the model gradually converges during the training process.}
    \label{fig:lossl}
\end{figure}

The training loss is shown in Fig. \ref{fig:lossl}. All experiments were conducted on an NVIDIA 4090 with an initial learning rate of 2e-4. We maintained a batch size of 8 and a sequence length limit of 2048 tokens. Each configuration ran for 2000 steps. Evaluation combined text matching metrics (BLEU, ROUGE) with formal syntax checking and functional simulation verification. The optimal configuration (r=16, all target layers) achieved 97\% functional accuracy on the test set, demonstrating significant improvement over baselines.
\begin{figure}
    \centering
    \includegraphics[width=1\linewidth]{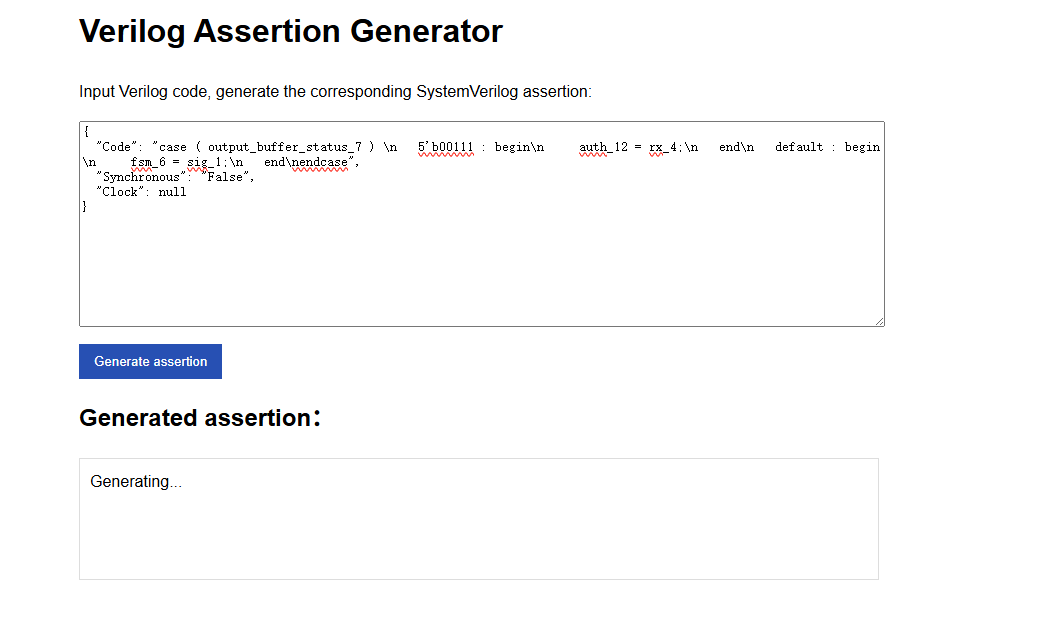}
    \caption{After entering the Verilog code, click on "Generate Assertion" and wait for the model to respond.}
    \label{fig:verlilog}
\end{figure}

\begin{figure}
    \centering
    \includegraphics[width=1\linewidth]{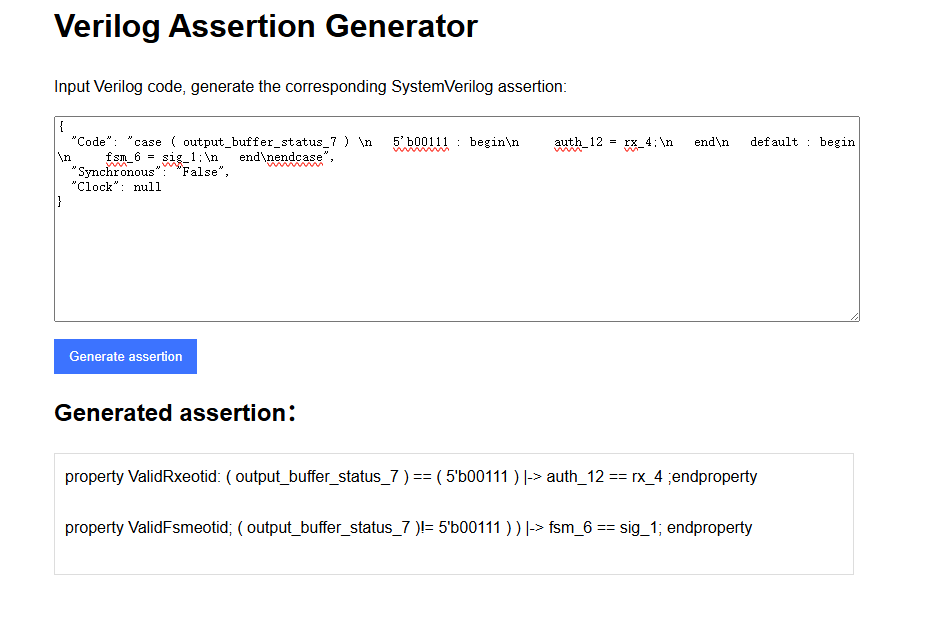}
    \caption{Model response completed, generating assertions corresponding to the Verilog code.}
    \label{verlilog2}
\end{figure}

\subsubsection{UI interface deployment}
To simplify the deployment process of our model training, we developed an easy-to-operate visualization interface that can automatically generate assertions from Verilog code. This tool enables users to easily upload their Verilog code, after which the system will perform a quick parsing and generate the corresponding assertion logic. The interface aims to enhance efficiency and usability, eliminating complex configuration requirements and achieving a seamless end-to-end workflow from code input to assertion generation. A specific demonstration is shown in the Fig. \ref{fig:verlilog} and Fig. \ref{verlilog2}.

\section{Discussion}
\subsection{The issue of inconsistent properties in the assertion}
\begin{figure}[h]
    \centering
    \includegraphics[width=\linewidth]{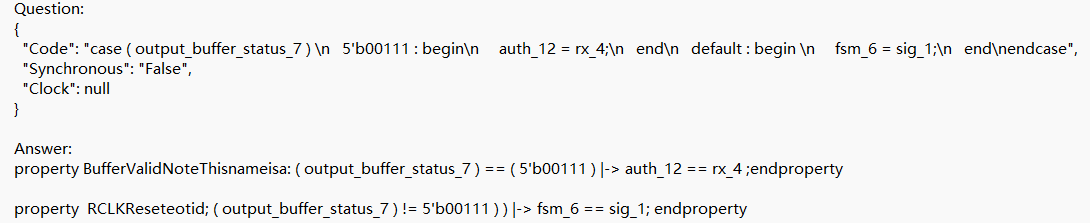}
    \caption{The "question" is followed by Verilog code, and the "answer" is the corresponding assertion. The above is the assertion corresponding to the code generated by our model.}
    \label{prediction}
\end{figure}

\begin{figure}[h]
    \centering
    \includegraphics[width=\linewidth]{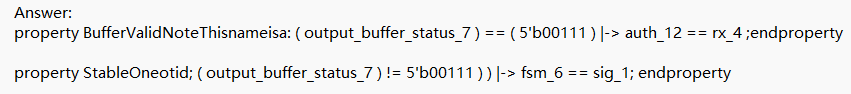}
    \caption{The above assertion is the true assertion within the dataset.}
    \label{label}
\end{figure}

As shown in Fig. \ref{prediction}, our model generated corresponding assertions based on the Verilog code. However, in fact, these assertions are different from those of the labels in the dataset, as shown in the Fig. \ref{label}. Their properties are not the same. But this does not affect the functionality of the assertions. This is just a custom attribute name. Therefore, our method of evaluating the accuracy rate only compares the content after the property, and does not require the property to be consistent.

\subsection{Dataset Simplicity and Future Directions}
The extremely high accuracy demonstrated in this study is due to the overly simplistic structure and content of the dataset. This dataset mainly focuses on a series of limited hardware verification tasks. However, as the complexity of these tasks increases - for example, by introducing more diverse hardware functions or multimodal inputs - merely relying on LoRA fine-tuning is insufficient.

To address these challenges, future research should delve into the collaborative integration of Direct Preference Optimization (DPO)~\cite{rafailov2023direct} and LoRA fine-tuning. DPO differs from traditional reinforcement learning (RL) methods such as PPO~\cite{schulman2017proximal, pan2023pattern, zhao2018reinforcement}, which rely on complex reward functions and iterative policy optimization. Instead, DPO reimagines the pattern of preference optimization as a more stable and computationally efficient supervised learning process. This innovative approach circumvents the severe instability problems faced by traditional RL techniques while reducing the high computational costs. Therefore, a framework that combines DPO and LoRA has the potential to achieve higher robustness and adaptability, enabling the model to handle increasingly complex and diverse verification scenarios.

\section{Conclusion}
This paper presents an efficient method for automatically generating assertions in hardware verification based on Verilog code. This method utilizes the LoRA-based fine-tuning technology within the Unsloth platform. Our approach significantly reduces the computational cost when extracting assertions from HDL code while maintaining high accuracy and generality. Experimental results show that this method can generate assertions that are both grammatically correct and semantically meaningful. Compared with traditional techniques, our method performs better and does not require a large amount of resource demand for extensive parameter fine-tuning. These results confirm the feasibility of lightweight optimization strategies in the hardware domain and provide a practical deployment method for resource-constrained environments.

Looking to the future, this research lays the foundation for more intelligent and adaptive tools in the field of automated testing and verification, especially in hardware-centric environments. Future extended versions may adopt reinforcement learning or few-shot learning models to further improve the quality and adaptability of assertions.

For instance, the framework based on DPO can be utilized to implement reinforcement learning. By integrating reward functions such as grammatical validity (e.g., SMT verification), semantic correctness, and mutation coverage, the improvement rate of fault detection can be optimized to 15\% - 20\%. Moreover, AdapterFusion \cite{pfeiffer2020adapterfusion} can be incorporated to achieve few-shot learning. By using only approximately 500 assertion templates during training, 98\% of the data requirements can be reduced, while still maintaining a performance of up to 90\%. Through this approach, we can further reduce the required training data and achieve fine-tuning for more scenarios.

\section{Contribution}
Yi Zhong developed the AutoAssert 1: a LoRA fine-tuned LLM model for assertion generation, Hongchao Liu did the data preprocessing  and Dr. Di Zhao directed the entire project.
\section{Acknowledgment}
We are grateful to Ms. Dantong Liu  from ARM and Mr. Xu Zhang and Mr. Hong Chen from Intel Labs China for their helpful discussions. We also thank Mr.  Yuan Gao for providing the Gitee code repository of AutoAssert1.


\end{document}